\documentclass[12pt,english]{article}

\usepackage[T1]{fontenc}
\usepackage[utf8]{inputenc}
\usepackage{geometry}
\usepackage{float}
\usepackage{amsmath}
\usepackage{graphicx}
\usepackage{setspace}
\usepackage{amssymb}
\usepackage{times}
\usepackage{indentfirst}
\makeatletter

\providecommand{\tabularnewline}{\\}

\newcommand{\lyxaddress}[1]{
	\par {\raggedright #1
	\vspace{1.4em}
	\noindent\par}
}

\usepackage{slashed}
\usepackage{hyperref}
\usepackage{graphicx}
\usepackage[numbers,sort&compress]{natbib}
\date{}

\hypersetup{colorlinks=true,citecolor=blue,linkcolor=red}

\makeatother

\usepackage{babel}
\begin{document}
\title{Traversable Thin-shell Wormhole in the 4D Einstein-Gauss-Bonnet Theory}
\author{Cheng-Yong Zhang$^{\ast1}$, Chao Niu$^{\ast1}$, Wei-Liang Qian$^{\ast3,4,5}$, Xiaobao Wang\thanks{zhangcy@email.jnu.edu.cn, niuchaophy@gmail.com, wlqian@usp.br, bao@mail.bnu.edu.cn } $^{2}$,
	Peng Liu\thanks{phylp@email.jnu.edu.cn (corresponding author)}	$^{1}$ }
\maketitle

\lyxaddress{\begin{center}

\textit{1. Department of Physics and Siyuan Laboratory, Jinan University, Guangzhou 510632, China}\\
\textit{2. College of Science, Beijing Information Science and Technology University, Beijing 100192, China }\\
\textit{3. Escola de Engenharia de Lorena, Universidade de S\~ao Paulo, 12602-810, Lorena, SP, Brazil}\\
\textit{4 Faculdade de Engenharia de Guaratinguet\'a, Universidade Estadual Paulista, 12516-410, Guaratinguet\'a, SP, Brazil}\\
\textit{5. Center for Gravitation and Cosmology, College of Physical Science and Technology, Yangzhou University, Yangzhou 225009, China}
\par\end{center}}

\begin{abstract}
This work investigates the spherically symmetric thin-shell wormhole solutions in four-dimensional Einstein-Gauss-Bonnet theory and explores their stabilities under radial, linear perturbations. 
These solutions are typically traversable and characterized by a thin-shell throat in accordance with Israel's junction conditions.
In asymptotically flat and AdS spacetimes with a negative Gauss-Bonnet coupling constant, stable neutral wormholes are encountered when the magnitude of the coupling constant becomes significant. 
The throats of such wormholes are sustained by ordinary matter and possess finite radii. 
In asymptotically dS spacetimes, no stable neutral wormhole featuring ordinary matter is observed. 
On the other hand, for positive Gauss-Bonnet coupling constant, stable thin-shell wormhole solutions can be established when the throats are exclusively supported by exotic matter. 
Moreover, stable charged wormholes comprised of ordinary matter are found universally in the asymptotically flat, AdS, and dS spacetimes. 
Unlike their neutral counterparts, the throat radii of such charged wormholes can be arbitrarily small. 
However, as the charge becomes more significant, such solutions only remain stable when supported by exotic matter.

\end{abstract}

\section{Introduction}\label{section1}

As a viable solution to Einstein's equation, a traversable wormhole is an intriguing object that connects two distinct regions of the universe via a throat.
The latter serves as a bridge allowing passage from one universe to the other. 
Modern wormhole physics was pioneered by Morris and Thorne~\cite{Morris1988}, where smooth traversable wormholes were proposed and comprised of exotic matter that violates the null energy condition (NEC)~\cite{Hochberg1998}. 
Among other alternatives, a specific class of traversable wormholes introduced by Visser~\cite{Visser1989}, referred to as thin-shell wormhole, is particularly interesting. 
Such a wormhole is constructed by gluing two spacetimes to form a geodesically complete manifold by placing a thin shell at the junction interface. 
Typically, the resultant wormhole metric is not smooth at its throat, which somewhat minimizes the usage of exotic matter. 
Nevertheless, in the framework of general relativity, the stable thin-shell wormholes violate the weak energy condition (WEC)~\cite{Poisson1995}. 
In this regard, many attempts have been made in alternative theories of gravity seeking traversable wormholes comprised of ordinary matter~\cite{Eiroa2005}.

In higher dimensional Einstein-Gauss-Bonnet (EGB) gravity, the Gauss-Bonnet (GB) term plays an effective role of negative energy density.
Subsequently, it allows for traversable thin-shell wormholes to be supported by ordinary matter~\cite{Gravanis2007, Simeone2005, Mazharimousavi2010, Zangeneh2015}. 
However, as a Lovelock invariant, the GB term is dynamically pertinent in spacetimes with the dimension $D\ge5$, as it becomes a total derivative in four-dimensional spacetimes. 
Recently, by rescaling the GB coupling constant $\alpha \to \alpha/(D-4)$, a four-dimensional EGB gravity theory was proposed in~\cite{Glavan20192019a}.
This work has triggered much attention in the literature regarding new black hole solutions and subsequent analyses~\cite{Fernandes2003,4DEGBSolutions, Cuyubamba2020,4DEGBShadow,4DEGBThermo,4DEGBOthers}. 
Although questions were raised about the validity of the underlying process, various remedies have been put forward, such as introducing an additional degree of freedom or breaking temporal diffeomorphism invariance~\cite{LU2020, Kobayashi2004}. 
The new black hole solutions, in turn, can also be adapted to these prescriptions. 
Notably, the properties of the spherically symmetric neutral black hole solutions in four-dimensional EGB gravity are somewhat distinct from those in higher dimensions.
For instance, in the former case, the gravitational force becomes repulsive near the object's center. 
In this regard, further exploring the properties of wormholes in this theoretical framework is meaningful.

The Morris-Thorne type wormholes were studied in~\cite{Jusufi2020}, where the flare-out conditions, embedding diagrams, and energy conditions were discussed.
The neutrual thin-shell wormholes were studied in \cite{Godani:2022jwz}. They focused on the cases with positive GB coupling constant and found that for the linear model, the thin-shell wormholes are unstable and filled with exotic matter. In contrast, for the nonlinear model, the solutions are stable and filled with non-exotic matter. 
In our work, we further investigate the thin-shell wormhole solutions and analyze their stability in the presence of linear radial perturbations. 
We investigate various scenarios associated with different metric configurations, such as the black hole's charge, the asymptotic behavior of the spacetime, and the GB coupling constant.
A rich structure regarding the stabilities of the underlying solutions is observed.

The remainder of the paper is organized as follows. 
Sec.~\ref{section2} gives a brief account of the 4D Einstein-Gauss-Bonnet gravity. 
The thin-shell wormhole solutions are derived in Sec.~\ref{section3}. 
In Sec~\ref{section4}, one analyzes the stability of the obtained thin-shell wormholes with respect to different configurations. 
The last section is devoted to further discussions and concluding remarks.

\section{The 4D Einstein-Gauss-Bonnet gravity}\label{section2}

We use the units $c=1,G=1$ throughout this paper. 
The action for the $D$ dimensional Gauss-Bonnet gravity in the presence of the Maxwell fields and a cosmological constant reads
\begin{equation}\label{eq:action0}
S=\frac{1}{16\pi}\int d^{D}x\sqrt{-g}\left[R+2\Lambda+\alpha\mathcal{G}-F_{\mu\nu}F^{\mu\nu}\right],
\end{equation}
where $g$ is the determinant of the metric, $R$ denotes the Ricci scalar, $\Lambda$ represents the cosmological constant, $\alpha$ gives the Gauss-Bonnet coupling constant, and the Gauss-Bonnet term $\mathcal{G}$ is defined as
\begin{equation}
\mathcal{G}=R^{2}-4R_{\mu\nu}R^{\mu\nu}+R_{\mu\nu\alpha\beta}R^{\mu\nu\alpha\beta},
\end{equation}
where $R_{\mu\nu}$ and $R_{\mu\nu\rho\sigma}$ are, respectively, the Ricci and Riemann tensors. 
The Maxwell tensor is given by $F_{\mu\nu}=\partial_{\mu}A_{\nu}-\partial_{\nu}A_{\mu}$, where $A_{\mu}$ is the gauge potential. 

To proceed, one introduces a counter-term into the action, which consists of the GB invariant of a conformally transformed metric $\tilde{g}_{\mu\nu}=e^{2\phi}g_{\mu\nu}$.
By taking the limit $D\to4$, the GB term is modified to give
\begin{equation}\label{eq:4dlim}
\lim_{D\to4}\frac{\int d^{D}x\sqrt{-g}\mathcal{G}-\int d^{D}x\sqrt{-\tilde{g}}\tilde{\mathcal{G}}}{D-4}=\int d^{4}\sqrt{-g}(4G^{\mu\nu}\partial_{\mu}\phi\partial_{\nu}\phi-\phi\mathcal{G}+4(\partial\phi)^{2}\square\phi+2(\partial\phi)^{4}).
\end{equation}
Here $G^{\mu \nu}$ is the Einstein tensor, and the factor $D-4$ comes from the parameter rescaling $\alpha \to \frac{\alpha}{D-4}$. 
Such a regularization procedure effectively removes the divergences in the four-dimensional limit and leads to a well-defined scalar-tensor theory of gravity. 
It is noted that the same theory could be obtained either by Kaluza-Klein reduction of the higher-dimensional EGB theory or by assuming a conformally invariant scalar field equation of motion~\cite{LU2020}. 
The above regularized 4D EGB theory possesses the following static spherically symmetric charged black hole solution with a cosmological constant, which is manifestly identical to that of the original theory~\cite{Fernandes2003}
\begin{equation}
ds^{2}=-f(r)dt^{2}+\frac{1}{f(r)}dr^{2}+r^{2}(d\theta^{2}+\sin^{2}\theta d\phi^{2}),\label{eq:metric}
\end{equation}
together with the gauge potential $A_{\mu}=-\frac{Q}{r}dt$, where $Q$ is the charge.
The metric function $f$ is given by
\begin{equation}
f(r)=1+\frac{r^{2}}{2\alpha}\left(1-\sqrt{1+4\alpha\left(\frac{2M}{r^{3}}-\frac{Q^{2}}{r^{4}}+\frac{\Lambda}{3}\right)}\right) ,
\end{equation}
where $M$ is the mass parameter.
We note that the solution (\ref{eq:metric}) can also be derived using the conformal anomaly \cite{Cai20019}. 
Also, a positive $\Lambda$ corresponds to an asymptotically dS spacetime, and a negative $\Lambda$ indicates an asymptotically AdS one. 
When $\alpha\to0$,   whether $\alpha$ is positive or negative, the Reissner-Nordstr\"om-(A)dS metric in general relativity can be recovered \cite{Fernandes2003}.

\section{The thin-shell wormhole solution}\label{section3}

In what follows, we construct a spherically symmetric thin-shell wormhole using the derived metric (\ref{eq:metric}). 
This is done by gluing together two copies of the spacetime region
\begin{equation}
\mathcal{M}_{1,2}=\{r_{1,2}\ge a\} ,
\end{equation}
which does not contain any horizon or singularity for $r>a$. 
Apparently, these two regions are geodesically incomplete manifolds with timelike boundaries $\Sigma_{1,2}=\{r_{1,2}=a\}$. 
Following Visser~\cite{Visser1989}, one may glue these two regions together through their boundaries by introducing a diffeomorphism using identical coordinates $\Sigma\equiv\Sigma_{1,2}$.
The boundary is a thin shell that plays the role of the wormhole's throat, such that the geodesic completeness holds for the new manifold $\mathcal{M}=\mathcal{M}_{1}\cup\mathcal{M}_{2}$. 
Also, as discussed below, it implies a surface energy-momentum tensor $S_{ij}$ localized at the wormhole's throat.
Based on the generalized Birkhoff theorem, the geometry for $r>a$ is still given by (\ref{eq:metric}).

On the timelike hypersurface $\Sigma$, we can define the line element
\begin{equation}
ds_{\Sigma}^{2}=-d\tau^{2}+a(\tau)^{2}(d\theta^{2}+\sin^{2}\theta d\phi^{2}),\label{eq:throat}
\end{equation}
where $\tau$ is the proper time measured by a comoving observer on the wormhole throat. 
The thin-shell geometry is assumed to be a function of $\tau$, which will be used as the temporal variable in studying the thin shell's mechanical stability. 
In general, the entire spacetime $\mathcal{M}$ is not smooth at the throat $\Sigma$, and an appropriate junction condition must be imposed. 
For EGB gravity of an arbitrary dimension, the generalized Darmois-Israel conditions determine the surface energy-momentum tensor $S_{ij}$ on $\Sigma$, which can be expressed as~\cite{Davis2002}
\begin{equation}\label{eq:junction}
S_{i}^{j}=-\frac{1}{8\pi}\left\langle K_{i}^{j}-K\delta_{i}^{j}\right\rangle -\frac{\alpha}{4\pi}\left\langle 3J_{i}^{j}-J\delta_{i}^{j}+2P_{imn}^{\ \ \ j}K^{mn}\right\rangle .
\end{equation}
Here the bracket means the jump (of the quantity inside the bracket) across the throat. 
As will be shown below, the surface energy-momentum tensor $S_{ij}$ satisfies the conservation equation $D_{i}S_{j}^{i}=0$, where $D_{i}$ is the covariant derivative on $\Sigma$.  
$K_{mn}$ is the extrinsic curvature of the throat, and its trace is denoted by $K=K_{i}^{i}$. 
The divergence-free parts of the Riemann tensor $P_{imnj}$ and the tensor $J_{ij}$ are given by
\begin{align}
P_{imnj}= & R_{imnj}+(R_{mn}g_{ij}-R_{mj}g_{in})-(R_{in}g_{mj}-R_{ij}g_{mn})+\frac{1}{2}R(g_{in}g_{mj}-g_{ij}g_{mn}), \\
J_{ij}=   & \frac{1}{3}(2KK_{im}K_{j}^{m}+K_{mn}K^{mn}K_{ij}-2K_{im}K^{mn}K_{nj}-K^{2}K_{ij}),
\end{align}
and the trace $J=J_{i}^{i}$. 
We note that the trace of the tensor $P_{imnj}$ is calculated with respect to the induced metric on the throat $\Sigma$. 
The generalized Darmois-Israel conditions come from the boundary term. 
On the other hand, however, the regularization procedure, i.e., the equation (\ref{eq:4dlim}), is irrelevant to the boundary.
As a result, the junction condition (\ref{eq:junction}) can be straightforwardly employed when $D=4$. 
In spherical symmetric spacetimes, the junction condition can also be derived with the help of the Gaussian normal coordinates in the vicinity of the throat, as shown in~\cite{Huang:2021bdm}. 
Nonetheless, the non-vanishing components of the surface energy-momentum tensor are found to be
\begin{align}
\sigma= & -\frac{\Delta}{4\pi}\left[\frac{2}{a}-\frac{4\alpha}{3a^{3}}\left(\Delta^{2}-3(1+\dot{a}^{2})\right)\right],\label{eq:sigma} \\
p= & \frac{1}{8\pi}\left(\frac{2\Delta}{a}+\frac{2l}{\Delta}-\frac{4\alpha}{3a^{2}}\left[3\Delta l-\frac{3l}{\Delta}(1+\dot{a}^{2})-\frac{\Delta^{3}}{a}-\frac{6\Delta}{a}\left(a\ddot{a}-\frac{1}{2}(1+\dot{a}^{2})\right)\right]\right) ,
\end{align}
where $\sigma=-S_{\tau}^{\tau}$ is the surface energy density, $p=S_{\theta}^{\theta}=S_{\phi}^{\phi}$ represents the transverse pressure exerted on the thin-shell,
$\Delta\equiv\sqrt{f(a)+\dot{a}^{2}}$, $l\equiv\ddot{a}+f'(a)/2$, where the overdot and prime denote derivatives with respect to $\tau$ and $a$, respectively. 
The above equations show that $S_{j}^{i}=\text{diag}(-\sigma,p,p)$. 
Specifically, the surface energy density and pressure satisfy the conservation equation,
\begin{equation}
0=\frac{d}{d\tau}(\sigma a^{2})+p\frac{d}{d\tau}a^{2}.\label{eq:conservation}
\end{equation}
Here, the first term on the r.h.s. of \eqref{eq:conservation} measures the change of the internal energy and the second term gives the work by the internal forces of the shell. 
For static thin-shell wormhole with radius $a_{0}$, we have
\begin{align}
\sigma_{0}= & -\frac{\sqrt{f(a_{0})}}{4\pi}\left[\frac{2}{a_{0}}-\frac{4\alpha}{3a_{0}^{3}}\left(f(a_{0})-3\right)\right],\label{eq:ED}  \\
p_{0}=      & \frac{\sqrt{f(a_{0})}}{8\pi}\left[\frac{2}{a_{0}}+\frac{f'(a_{0})}{f(a_{0})}-\frac{4\alpha}{a_{0}^{2}}\left(\frac{f'(a_{0})}{2}-\frac{f'(a_{0})}{2f(a_{0})}-\frac{f(a_{0})}{3a_{0}}+\frac{1}{a_{0}}\right)\right].\label{eq:P}
\end{align}

Before closing this section, we elaborate on the energy conditions for the static thin-shell wormhole. 
The WEC states that for any timelike vector $U^{\mu}$ there must be $T_{\mu\nu}U^{\mu}U^{\nu}\ge0$.
Also, the NEC requires that for any null vector $k^{\mu}$ there must be $T_{\mu\nu}k^{\mu}k^{\nu}\ge0$. 
In an orthogonal basis, the WEC implies $\rho\ge0,\rho+P_{i}\ge0$ and NEC reads $\rho+P_{i}\ge0$ where $i\in\{\theta,\phi\}$ where $\rho$ is the energy density and $P_{i}$ is the pressure along the $i$-th coordinate. 
In the case of the spherical thin-shell wormhole we considered here, the radial pressure $P_{r}=0$ and $\rho=\sigma_{0}\delta(r-a_{0})$. 
Therefore, the WEC and NEC coincide and simplify to $\sigma\ge0$. 
The matter violating the energy conditions is called exotic matter. 
In literature, the total amount of exotic matter has been suggested as an indicator of the physical viability of traversable wormholes~\cite{Nandi2004}. 
The most common choice of the total amount of the matter in the static wormhole is given by \cite{Visser2003}
\begin{equation}
\Omega=\int(\rho+P_{r})\sqrt{-g}drd\theta d\phi=4\pi a_{0}^{2}\sigma_{0} ,
\end{equation}
which is proportional to the energy density $\sigma_{0}$. 
According to (\ref{eq:ED}) and (\ref{eq:P}), we observe that when the wormhole radius $a_{0}$ tends to the event horizon in the original metric, the energy density approaches zero, and meanwhile, the pressure diverges unless $f'(a_{0})=0$. 
The latter condition corresponds to the extremal black holes, and the conclusion generally holds in the framework of general relativity.
For the 4D EGB gravity, however, when $\alpha=-a_{0}^{2}/2$, the pressure on the throat remains finite.

\section{Stability analysis of the thin-shell wormholes}\label{section4}

To study the stability of the thin-shell wormhole under a radial perturbation of the throat radius, we assume the following linear equation of state for the matter~\cite{Poisson1995},
\begin{equation}
p=p_{0}+\eta(\sigma-\sigma_{0}), \label{eq:perturb} 
\end{equation}
where $\sigma_{0}$ and $p_{0}$ are the energy density and pressure on the static thin-shell at $r=a_{0}$. 
For wormholes supported by ordinary matter, $\eta^{1/2}$ is the speed of sound of  matter. But there is no guarantee that $\eta^{1/2}$ is actually   the speed of   sound since we do not know the detailed microphysics for exotic matter \cite{Poisson1995}.  
Here we assume it to be an arbitrary constant due to the presence of exotic matter.
Using (\ref{eq:perturb}) and the conservation equation (\ref{eq:conservation}), we find
\begin{equation}
\sigma(a)=\frac{\sigma_{0}+p_{0}}{\eta+1}\left(\frac{a_{0}}{a}\right)^{2(\eta+1)}+\frac{\eta\sigma_{0}-p_{0}}{\eta+1}.\label{eq:sigmaPer}
\end{equation}

By comparing \eqref{eq:sigmaPer} against (\ref{eq:sigma}), one obtains the following one-dimensional equation of motion
\begin{equation}
\dot{a}^{2}+V(a)=0 ,
\end{equation}
where the potential $V$ for the equation of motion reads
\begin{equation}
V(a)=f(a)-\left[(\sqrt{A^{2}+B^{3}}-A)^{1/3}-\frac{B}{(\sqrt{A^{2}+B^{3}}-A)^{1/3}}\right]^{2},
\end{equation}
and
\begin{equation}
A=\frac{3\pi a^{3}}{4\alpha}\sigma(a),\ B=\frac{a^{2}}{4\alpha}+\frac{1-f(a)}{2},
\end{equation}
where $\sigma(a)$ is given by (\ref{eq:sigmaPer}). 

At the limit $\alpha\to0$, we have
\begin{align}
V= & f-\left[2\pi a\sigma(a)\right]^{2},
\end{align}
which is identical to what was found in \cite{Poisson1995}. 

Intuitively, the shape of the potential $V(a)$ governs the stability of the thin-shell wormhole under linear perturbation. 
It can be shown that both $V(a_{0})$ and $V'(a_{0})$ vanishes, and thus one has to explore the higher order terms.
The expansion around $a=a_{0}$ at the leading non-vanishing order gives
\begin{equation}
\dot{a}^{2}+\frac{1}{2}V''(a_{0})(a-a_{0})^{2}=0.
\end{equation}
Therefore, the requirement for stable equilibrium implies $V''(a_{0})>0$. 
In practice, the specific form of $V''(a_{0})$ is rather complicated, which will be studied in the following subsections.
We will analyze the stable region of the thin-shell wormholes in asymptotic flat, dS, and AdS spacetimes and for different metric configurations. 
Without loss of generality, hereafter, we assume $M=1$. 
The parameter space will be scrutinized in terms of the remaining metric parameters, particularly $\Lambda$, $Q$, $\alpha$, and $\eta$.

\subsection{Asymptotically flat spacetimes}
{\footnotesize{}}
\begin{figure}[t]
	\begin{centering}
		{\footnotesize{}}%
		\includegraphics[scale=0.6]{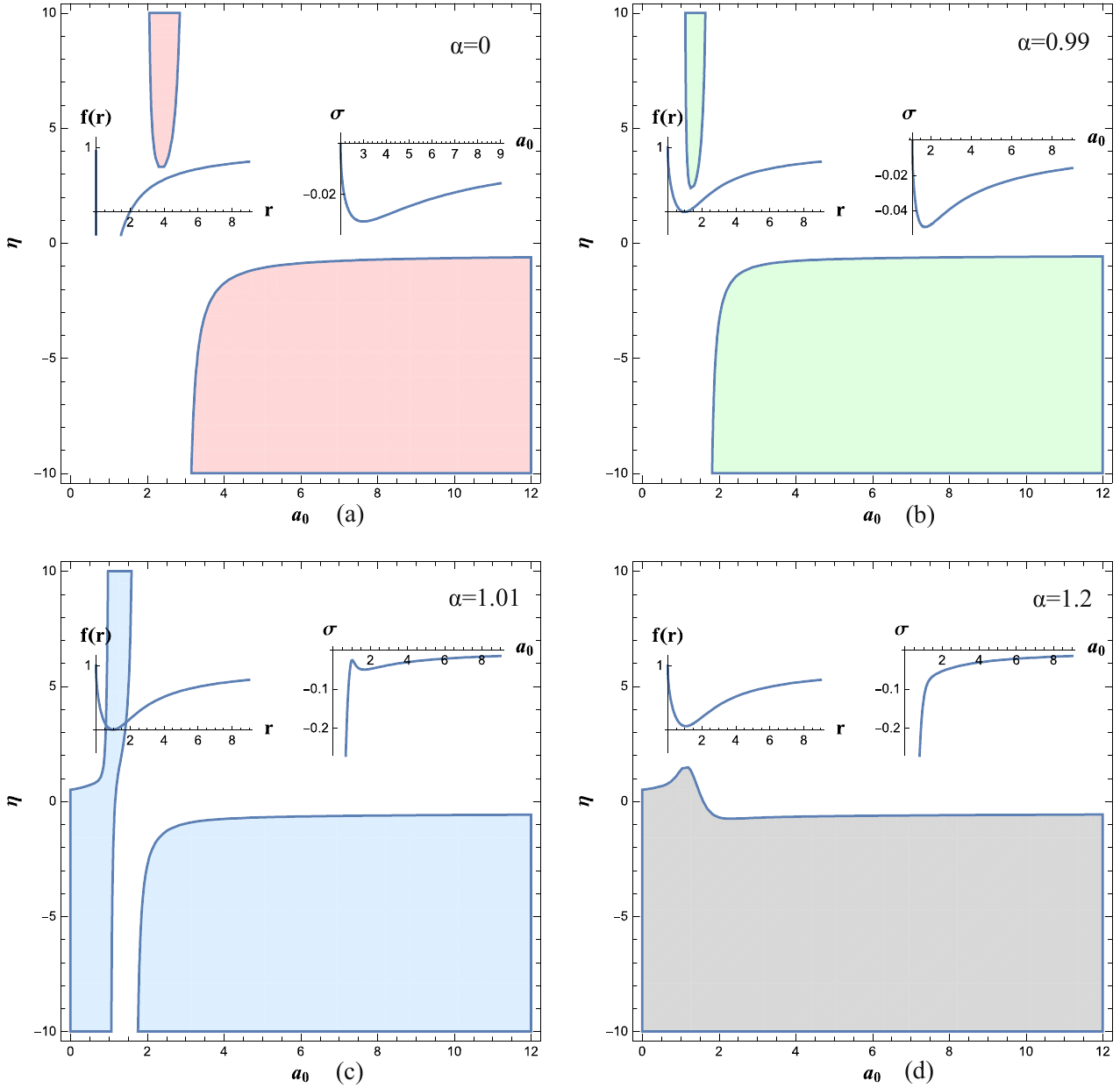}
		{\footnotesize\par}
		\par\end{centering}
	{\footnotesize{}\caption{\label{fig:FlatQ0A}{\small{}The stable regions of the thin-shell wormholes in 4D EGB gravity with $\Lambda=0$, $Q=0$, and $\alpha \ge 0$.
	The subfigures marked with (a,b,c,d)  give the results for $\alpha=0$, $0.99$, $1.01$, and $1.2$, respectively. 
	The corresponding metric function $f(r)$ and energy density $\sigma_{0}$ are presented in the insets.}}
	}{\footnotesize\par}
\end{figure}
{\footnotesize\par}

For asymptotic flat spacetime, the cosmological constant $\Lambda=0$ and the metric function
\begin{equation}
f(r)=1+\frac{r^{2}}{2\alpha}\left(1-\sqrt{1+4\alpha\left(\frac{2M}{r^{3}}-\frac{Q^{2}}{r^{4}}\right)}\right)
\end{equation}
has two roots
\begin{equation}
r_{\pm}=M\pm\sqrt{M^{2}-Q^{2}-\alpha}.
\end{equation}
It is worth noting that we are not limited by the constraint $M^{2}-Q^{2}-\alpha>0$. 
The condition $M^{2}-Q^{2}-\alpha>0$ is a requirement for the existence of a    black hole horizon, not for the existence of a traversable thin-shell wormhole. If $M^{2}-Q^{2}-\alpha>0$, we must guarantee that the traversable thin-shell wormhole throat radius $a_{0}$ is larger than the black hole horizon radius $r_{+}$. If $M^{2}-Q^{2}-\alpha<0$, there is no black hole horizon, and we must guarantee that the singularity is surrounded by the traversable thin-shell wormhole throat to ensure that the geometry is well-defined. 

\begin{figure}[H]
	\begin{centering}
		{\footnotesize{}}%
		\includegraphics[scale=0.6]{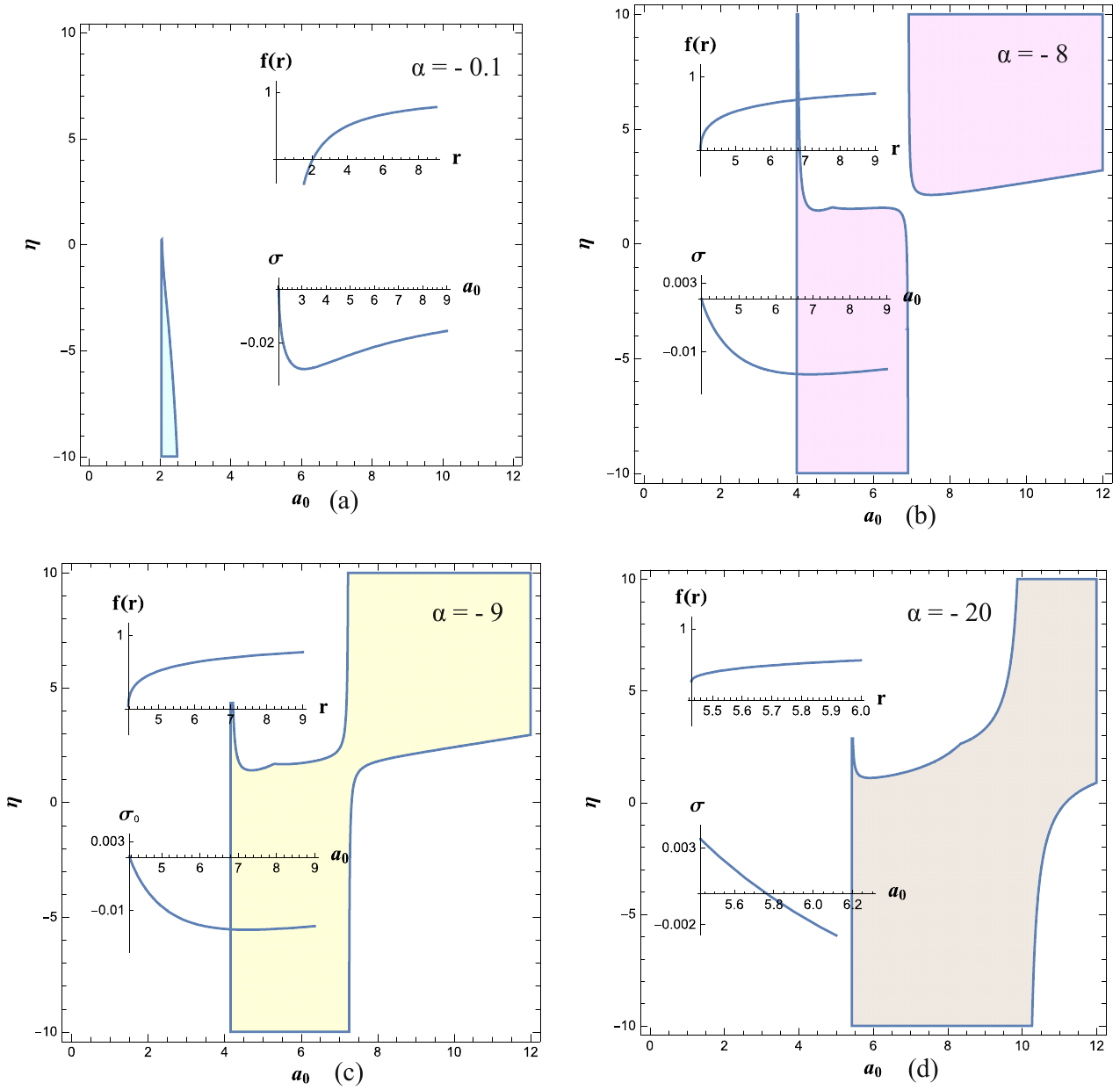}
		{\footnotesize\par}
		\par\end{centering}
	{\footnotesize{}\caption{\label{fig:FlatQ0Am}{\small{}The same as Fig.~\ref{fig:FlatQ0A} but for $\alpha < 0$. 
	The subfigures marked with (a,b,c,d)  give the results for $\alpha=-0.1$, $-8$, $-9$, and $-20$, respectively. }}
	}{\footnotesize\par}
\end{figure}

In Fig.~\ref{fig:FlatQ0A}, we show the stable region of the thin-shell wormhole in 4D EGB gravity with $Q=0$ and for various choices of $\alpha \ge 0$.
Fig. \ref{fig:FlatQ0Am} gives the corresponding results for various choices of $\alpha<0$. 
For positive but moderately small $\alpha$, it is observed that the radius of the wormhole's throat is bound from below.
As $\alpha$ increases, the size of the stable region mainly increases. 
However, as $\alpha$ further increases, the throat's radius of the thin-shell wormhole can be arbitrarily small, and the stable region only persists primarily for $\eta <0$.
The energy densities of the thin-shell wormholes are also presented. 
For $\alpha \ge 0$, the energy density is always found negative; therefore, the matter supporting the stable thin-shell wormhole is exotic.

{\footnotesize{}}
{\footnotesize\par}

The scenario is more interesting for negative $\alpha$. 
As $\alpha$ approaches $0_-$, the stable region suddenly shrinks.
In this case, the stable solutions are only feasible for $\eta<0$ and in a narrow range for the throat's radius, as shown in    Fig.~\ref{fig:FlatQ0Am}(a). 
As $\alpha$ further decreases, stable solutions with positive $\eta$ also become possible, provided that the throat's radius is large enough. 
When $\alpha$ further decreases, stable wormholes can be supported by ordinary matter with positive energy density, as shown in   Fig.~\ref{fig:FlatQ0Am}(d). 
It is noted that $\alpha$ is unbound from below. 
In other words, the stable wormholes supported by ordinary matter always persist for negative and significant $\alpha$.

\begin{figure}[H]
	\begin{centering}
		{\footnotesize{}}%
		\includegraphics[scale=0.56]{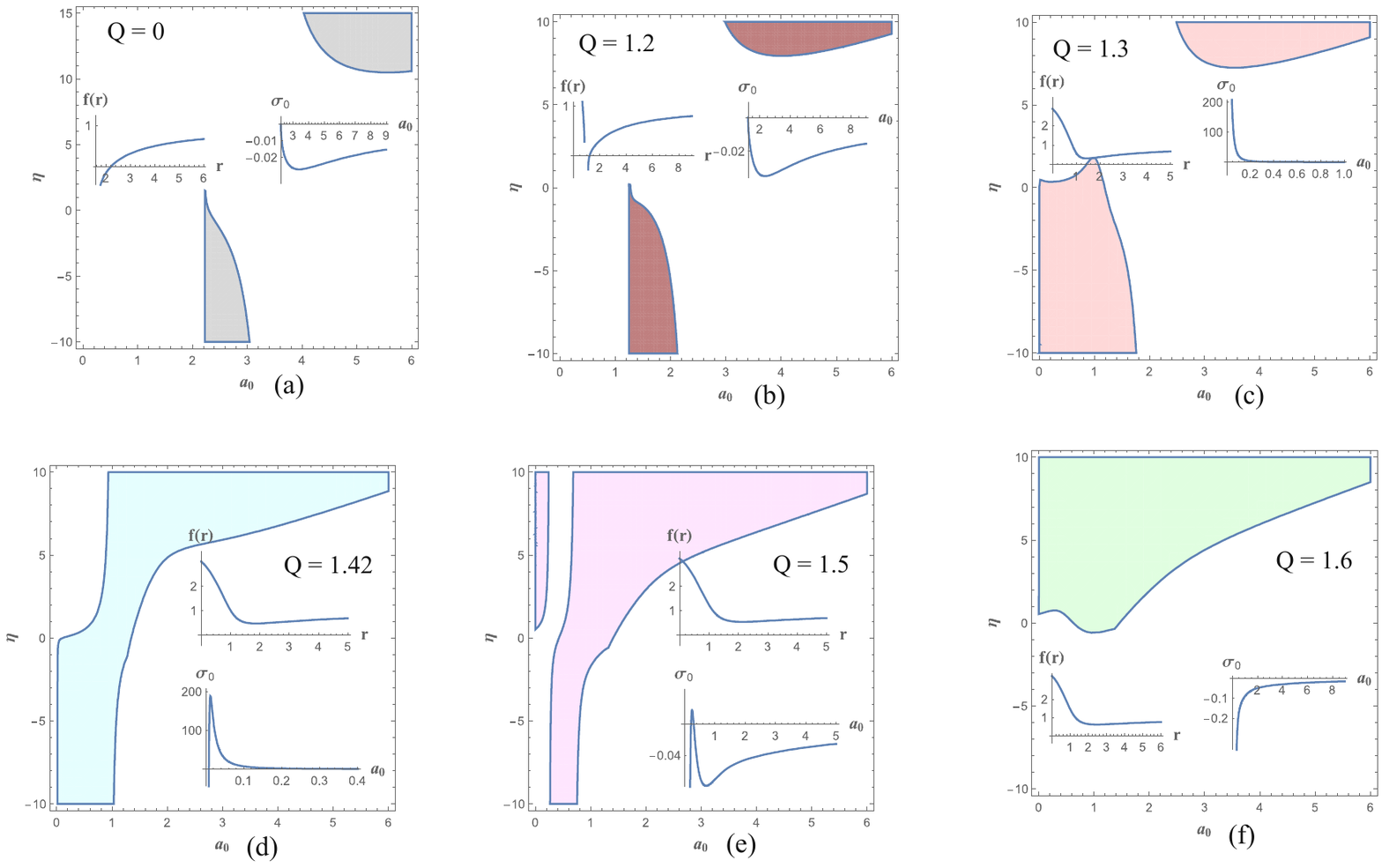}
		{\footnotesize\par}
		\par\end{centering}
	{\footnotesize{}\caption{\label{fig:FlatAlpha05m}{\small{}The stable regions of the thin-shell wormholes in 4D EGB gravity with $\Lambda=0$ and $\alpha =-0.5$ but for different values of charge $Q$.
	The subfigures marked with (a,b,c,d,e,f)  give the results for $Q=0$, $1.2$, $1.3$, $1.42$, $1.5$, and $1.6$, respectively. 
	The corresponding metric function $f(r)$ and energy density $\sigma_{0}$ are presented in the insets.}}
	}{\footnotesize\par}
\end{figure}

Now let us study the effect of the charge $Q$.
As shown in Fig.~\ref{fig:FlatAlpha05m}, the matter supporting the stable wormhole is exotic when $Q$ is small. 
However, as $Q$ increases, the matter supporting the stable wormhole can be ordinary, provided the wormhole's throat is not too large. 
However, when $Q$ becomes large enough, again, only exotic matter can support stable wormhole solutions. 
Regarding the stable solutions, it is noted that $Q$ is unbound from above.

{\footnotesize{}}
{\footnotesize\par}

\subsection{Asymptotically dS spacetimes}

\begin{figure}[H]
	\begin{centering}
		{\footnotesize{}}%
		\includegraphics[scale=0.6]{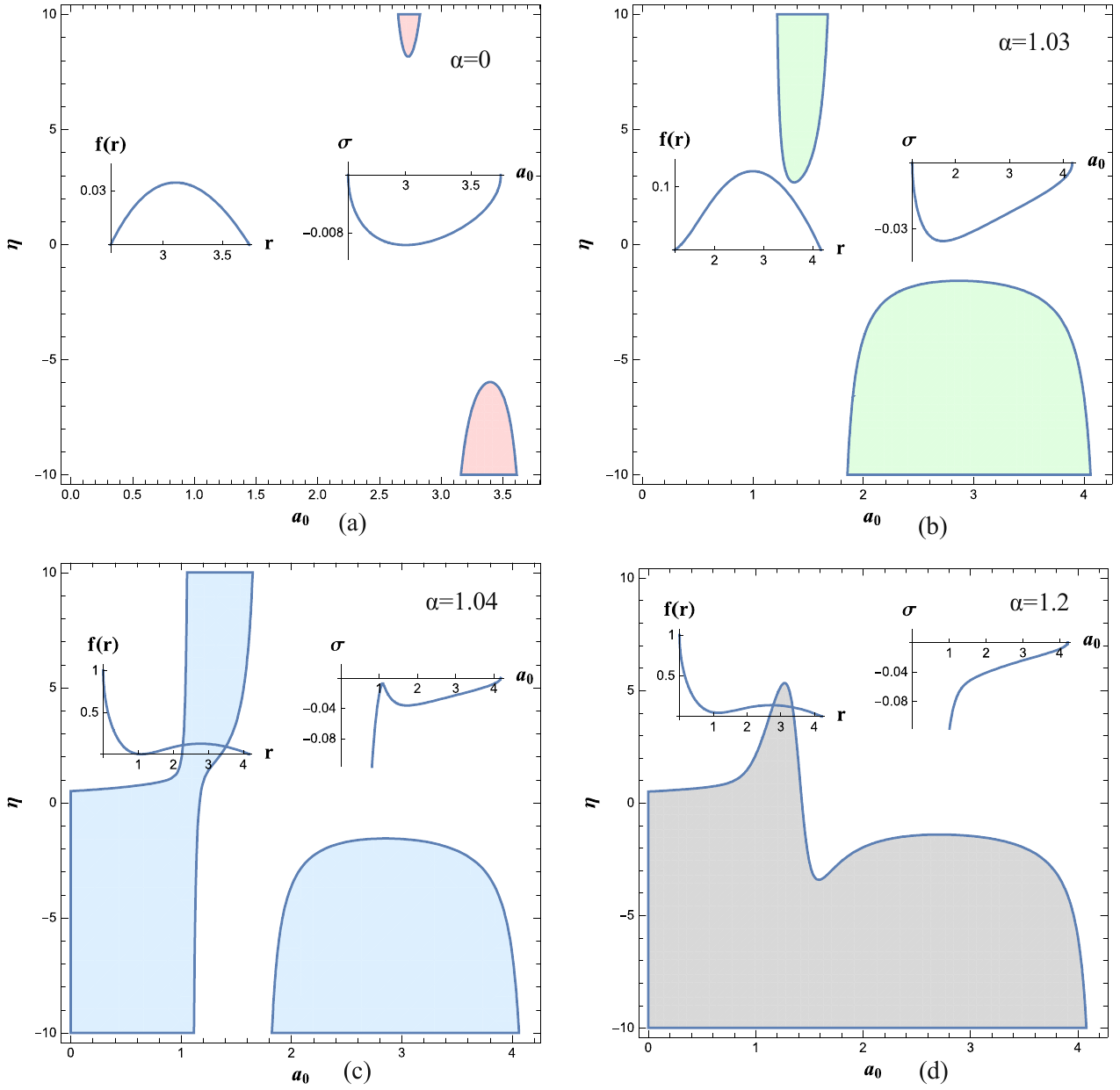}
		{\footnotesize\par}
		\par\end{centering}
	{\footnotesize{}\caption{\label{fig:dSQ0A}{\small{}The stable regions of the thin-shell wormholes in 4D EGB gravity with $\Lambda=0.1$, $Q=0$, and $\alpha \ge 0$.
	The subfigures marked with (a,b,c,d)  give the results for $\alpha=0$, $1.03$, $1.04$, and $1.2$, respectively. 
	The corresponding metric function $f(r)$ and energy density $\sigma_{0}$ are presented in the insets.}}
	}{\footnotesize\par}
\end{figure}

\begin{figure}[H]
	\begin{centering}
		{\footnotesize{}}%
		\includegraphics[scale=0.6]{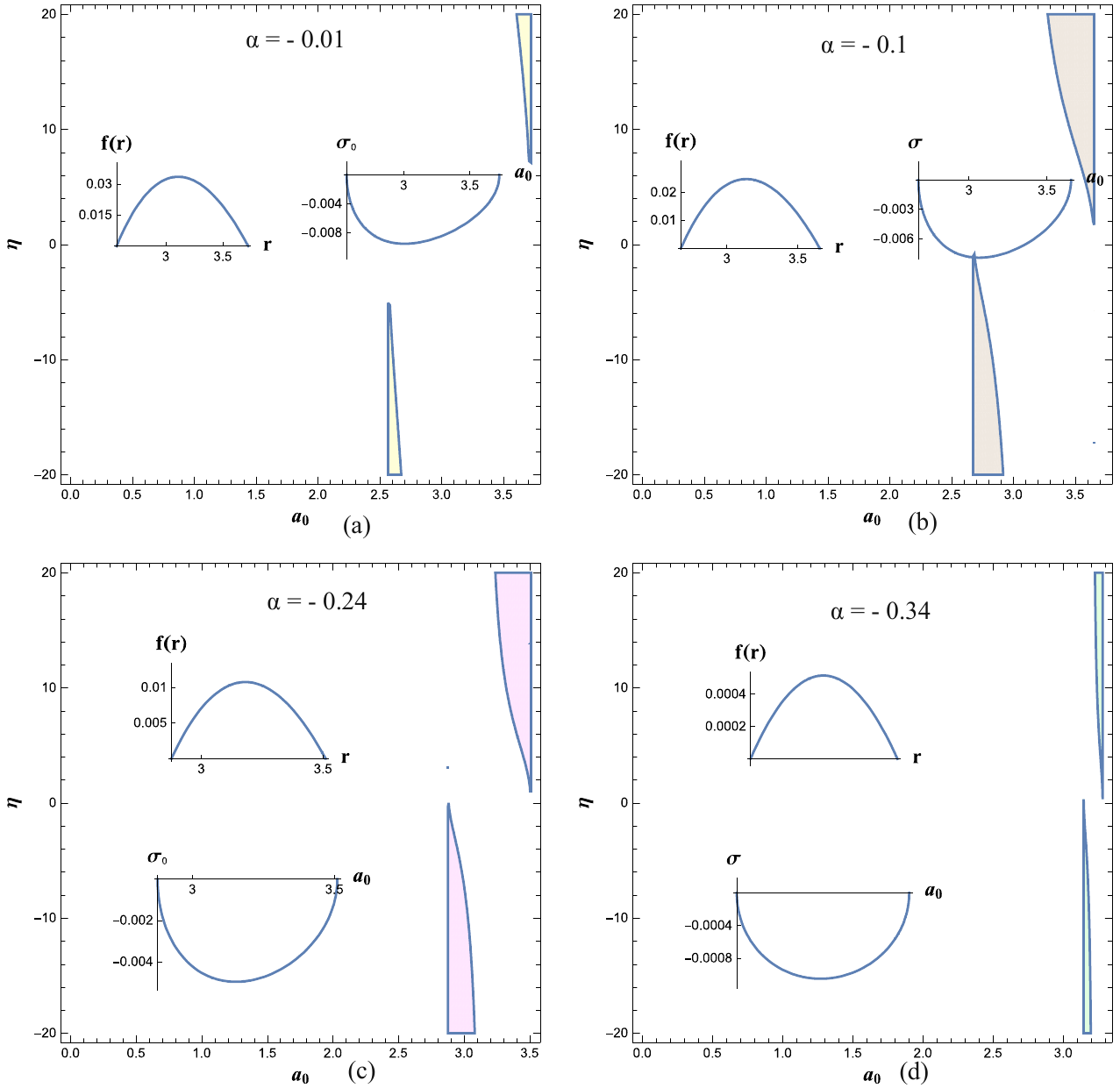}
		{\footnotesize\par}
		\par\end{centering}
	{\footnotesize{}\caption{\label{fig:dSQ0Am}{\small{}The same as Fig.~\ref{fig:dSQ0A} but for $\alpha < 0$. 
	The subfigures marked with (a,b,c,d)  give the results for $\alpha=-0.01$, $-0.1$, $-0.24$, and $-0.34$, respectively. }}
	}{\footnotesize\par}
\end{figure}

Now, we turn to explore the wormholes in asymptotically dS spacetimes. 
For convenience, we assume the cosmological constant to be $\Lambda=0.1$ for the numerical calculations carried out in this subsection. 
The stable regions of the neutral thin-shell wormholes in 4D EGB-dS wormholes are presented in Fig.~\ref{fig:dSQ0A} and~\ref{fig:dSQ0Am} for, respectively, $\alpha\ge 0$ and $\alpha<0$. 
For $\alpha>0$, the stable regions possess essentially a similar feature as observed in the asymptotically flat spacetimes.
The only difference is that the throat's radius is constrained by the cosmological horizon. 
Again, these neutral, stable, thin-shell wormholes are only supported by exotic matter.

{\footnotesize{}}
{\footnotesize\par}

On the other hand, the cases for $\alpha<0$ are found to be somewhat different from those in asymptotically flat spacetimes. 
As demonstrated in Fig.~\ref{fig:dSQ0Am}, as $\alpha$ decreases with increasing magnitude, the size of the stable region increases and then shrinks. 
It is noted that the possible value of the GB coupling constant $\alpha$ is bound from below by a critical value $\alpha>\alpha_{c}=-0.345$. 
For $\alpha <\alpha_c$, no stable thin-shell wormhole is encountered as the size of the stable region vanishes.

	{\footnotesize{}}
{\footnotesize\par}

\begin{figure}[H]
	\begin{centering}
		{\footnotesize{}}%
		\includegraphics[scale=0.66]{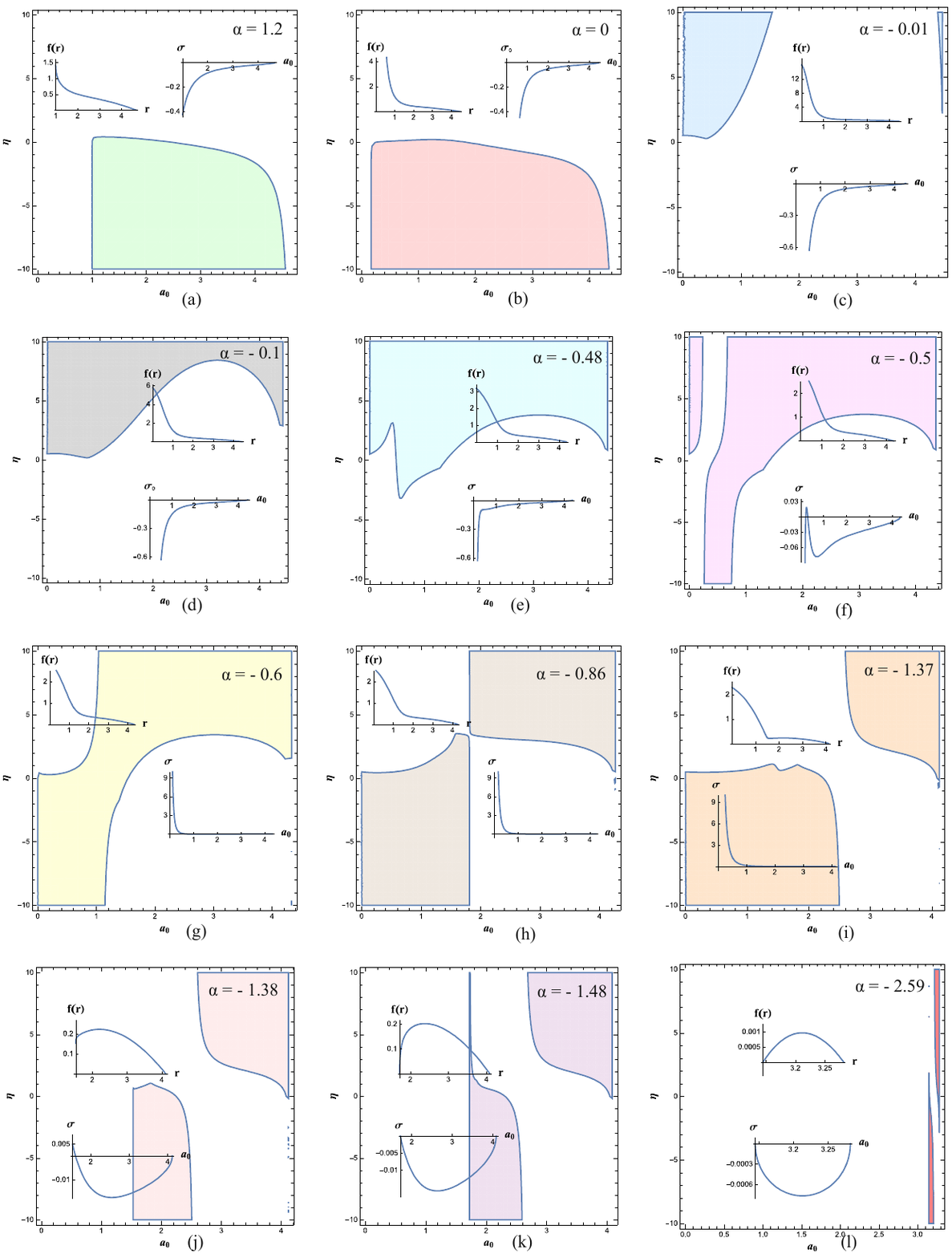}
		{\footnotesize{}.}{\footnotesize\par}
		\par\end{centering}
	{\footnotesize{}\caption{\label{fig:dSQ06A}{\small{}The same as Fig.~\ref{fig:dSQ0A} but for $Q=0.6$ and different values of $\alpha$. The subfigures marked with (a,b,c,d,e,f,g,h,i,j,k,l)  correspond to $\alpha=1.2$, $0$, $-0.01$ (first row), $-0.1$, $-0.48$, $-0.5$ (second row), $-0.6,-0.86,-1.37$ (third row), and $-1.38$, $-1.48$, $-2.59$ (last row), respectively.}}
	}{\footnotesize\par}
\end{figure}

In Fig.~\ref{fig:dSQ06A}, we present the scenarios for charged wormholes with $Q=0.6$.
For positive $\alpha$, the stable region shrinks as $\alpha$ increases. 
These stable solutions are all supported by exotic matter with $\eta<0$, as indicated in   Fig.~\ref{fig:dSQ06A}(a,b). 
When $\alpha$ becomes negative but with moderate magnitude, the stable region switches to $\eta>0$. 
For the interval $-0.5>\alpha>-1.38$, stable charged wormholes are supported by ordinary matter featuring positive energy density and positive  $\eta^{1/2}$. 
However, as $\alpha$ decreases further, the stable region shrinks, and the only viable solutions are supported by exotic matter.

{\footnotesize{}}
{\footnotesize\par}

\subsection{Asymptotically AdS spacetimes}

\begin{figure}[H]
	\begin{centering}
		{\footnotesize{}}%
		\begin{tabular}{cc}
			{\footnotesize{}\includegraphics[scale=0.75]{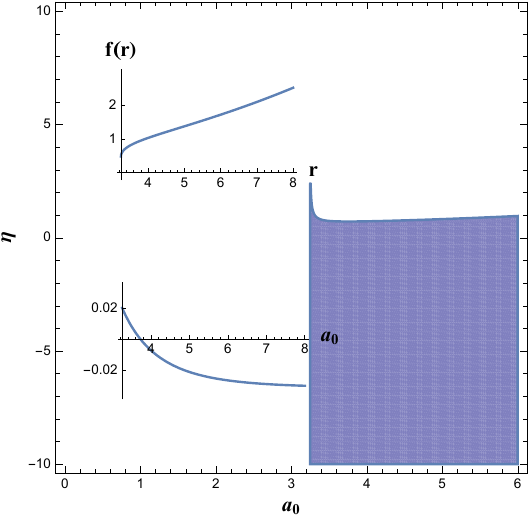}} & {\footnotesize{}\includegraphics[scale=0.8]{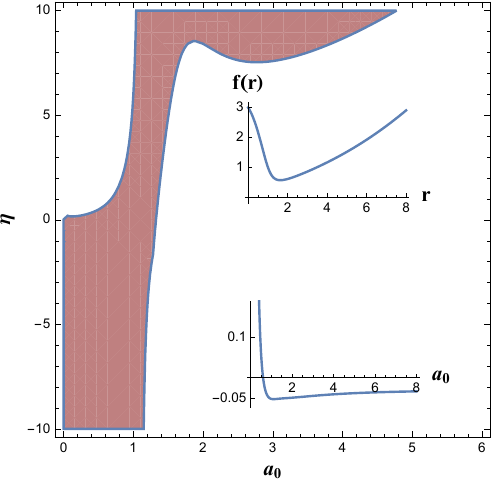}}\tabularnewline
		\end{tabular}{\footnotesize\par}
		\par\end{centering}
	{\footnotesize{}\caption{\label{fig:AdSQA}{\small{}The stable regions of the thin-shell wormholes in 4D EGB gravity with $\Lambda=-0.1$.
	The left and right panels give the results for $Q=0$, $\alpha=-10$, and $Q=1.4$, $\alpha=-0.5$, respectively. 
	The corresponding metric function $f(r)$ and energy density $\sigma_{0}$ are presented in the insets.}}
	}{\footnotesize\par}
\end{figure}

We study the thin-shell wormhole solutions in asymptotically AdS spacetimes in the subsection.
The properties of the stable regions related to the thin-shell wormholes in AdS spacetime are mainly reminiscent of those in asymptotically flat spacetimes. 
Here, we present two typical cases in Fig.~\ref{fig:AdSQA}. 
For $\alpha>0$, the stable thin-shell wormholes are exclusively supported by exotic matter. 
For $\alpha<0$, on the other hand, the stable solutions can be supported by ordinary matter. 
The throat radius is typically more significant for a neutral wormhole and bound from below. 
While for charged wormholes, the throat radius can be arbitrarily small.

{\footnotesize{}}
{\footnotesize\par}

\section{Concluding remarks}\label{section5}

To summarize, we constructed the thin-shell wormholes in the 4D Einstein-Gauss-Bonnet gravity minimally coupled to the electromagnetic field in asymptotically flat, dS, and AdS spacetimes.
The linear stabilities of the solutions under radial perturbations were investigated, which revealed a rather rich structure regarding different spacetime configurations.
For positive Gauss-Bonnet coupling constant, stable thin-shell wormholes can be supported only by exotic matter in flat, dS, and AdS spacetime.  These results are consistent with those found in \cite{Godani:2022jwz} which claimed that for the linear model, the neutral thin-shell wormholes in asymptotically flat  spacetimes are unstable and filled with exotic matter  when $\alpha>0$.  The hopes for obtaining stable thin-shell wormholes with ordinary matter when $\alpha>0$ have already been dashed \cite{Simeone2005, Mazharimousavi2010, Zangeneh2015}. . 

On the other hand, different scenarios are encountered for negative GB coupling constant.
In asymptotically flat and AdS spacetimes, stable neutral wormhole solutions supported by the ordinary matter are observed, which feature finite but large throat radii.
On the other hand, a charged wormhole supported by ordinary matter may possess either a large throat radius or an arbitrarily small one. 
The GB coupling constant is not bound from below for the latter case. 
However, the charge of the wormhole is found to be constrained in a small window. 
Otherwise, it can only be supported by exotic matter.
In asymptotic dS spacetimes, there is no stable neutral wormhole solution supported by ordinary matter. 
Again, the GB coupling constant is bound from below. 
The charged ones can be stable and supported by ordinary matter only when the GB coupling constant is negative with moderate magnitude.
It was pointed out that the energy condition of the Morris-Thorne wormhole is violated at the wormhole's throat for positive GB coupling constant~\cite{Jusufi2020}. 
However, it was expected that for a negative GB coupling constant, the energy condition could be satisfied~\cite {Zangeneh2015}.
The results obtained in the present study are primarily consistent with this observation.

\section*{Acknowledgments}

We thank Yu Tian, Yong-Ming Huang,  Minyong Guo, and Peng-Cheng Li for their helpful discussions. 
Peng Liu would like to thank Yun-Ha Zha for her kind encouragement during this work. 
Peng Liu is supported by the Natural Science Foundation of China under Grant No. 11847055 and 11905083. 
Chao Niu is supported by the Natural Science Foundation of China under Grant No. 11805083. 
C. Y. Zhang is supported by the Natural Science Foundation of China under Grant No. 11947067, 12005077, and Guangdong Basic and Applied Basic Research Foundation (2021A1515012374).
We also acknowledge the financial support from Brazilian agencies 
Funda\c{c}\~ao de Amparo \`a Pesquisa do Estado de S\~ao Paulo (FAPESP),
Funda\c{c}\~ao de Amparo \`a Pesquisa do Estado do Rio de Janeiro (FAPERJ),
Conselho Nacional de Desenvolvimento Cient\'{\i}fico e Tecnol\'ogico (CNPq),
Coordena\c{c}\~ao de Aperfei\c{c}oamento de Pessoal de N\'ivel Superior (CAPES).

\end{document}